# Broadband acoustic trapping of a particle by a soft plate with a periodic deep grating

Hailong He, Shiliang Ouyang, Zhaojian He[aa)], Ke Deng[a)], and Heping Zhao

*Department of Physics, Jishou University, Jishou 416000, Hunan, China*

We investigated the acoustic radiation force (ARF) acting on a cylindrical brass particle near an acoustically soft plate patterned with a periodic deep grating. The existence of a negative ARF by which the particle can be pulled towards the sound source is confirmed. In addition, the bandwidth for negative ARF in this soft-plate system is found to be considerably broader than in the stiff-plate systems typically used in previous studies. It is further demonstrated by field distribution analysis that the negative ARF is caused by the gradient force induced by the gradient vortex velocity field near the surface, which stems from the collective resonance excitation of the antisymmetric coupling of Scholte surface waves in the thin plate. The effects of particle location and size on the ARF were also investigated in detail. The negative ARF has potential use in applications requiring particle manipulation using acoustic waves.

## I. INTRODUCTION

When impinging on objects, acoustic waves can produce acoustic radiation forces (ARFs) as a result of momentum exchange between the object and the incident field [1, 2]. In most situations, such forces are positive because acoustic waves usually push objects in the direction of propagation away from the sound source. However, the negative ARF (acoustic waves pulling objects towards the sound source) has attracted attraction recently because of its potential application in acoustic manipulation [3–5] and acoustic levitation [6] of particles, thereby offering a contactless and noninvasive tool for many applications, such as cell sorting and collection in biological engineering [7,8]. It has been demonstrated that negative ARFs can be obtained by using nondiffracting acoustic beams, such as Bessel beams [9–14] or crossed plane waves [15, 16]. In these cases, there are acoustic force gradients in the standing waves, with the direction of the force being either toward the pressure nodes or antinodes, which may be in the direction of the sound source, thus giving rise to negative ARFs. More complicated manipulations including sharp bending of microparticles by acoustic half-Bessel beams have also been demonstrated very recently [17].

---

[a) a)]Authors to whom correspondence should be addressed. Email: hezj@jsu.edu.cn and dengke@jsu.edu.cn

In the past few years, artificial acoustic structured plates have attracted great interest because of their exotic intrinsic modes [18–20]. In particular, resonance excitation of antisymmetric Lamb waves in an acoustically stiff plate can lead to greatly enhanced acoustic transmissions [21]. Unique near-field vortex patterns were confirmed in such exotic wave responses, which provide a rich diversity of interactions between waves and objects. Acoustic trapping of small particles in such vortex fields near a periodically structured brass plate immersed in water has been reported [3, 4]. In contrast to conventional acoustic trapping which uses standing waves, trapping of subwavelength (relative to the wavelength of sound in water) particles can be realized for this plate system because the wavelength of excited plate modes is generally much shorter than that of ultrasound in water. In addition, it was observed [4] that acoustic resonance produces strong acoustic fields near the plate and greatly enhances ARFs without requiring a higher input power. A similar result was reported in Ref. [22], in which ARFs exerted on a rigid wall were found to be markedly amplified by resonance coupling with adjoining metamaterial slabs. Moreover, a recent study investigating interactions between a pair of stiff structured plates provided evidence of an attractive force caused by near-field coupling [23].

In this paper, ARFs acting on a cylindrical brass particle are investigated using an acoustically soft plate patterned with a periodic deep grating. It is confirmed that the ARF could turn negative when the particle is immersed in the near-field of the vortex around the plate. In addition, the bandwidth for the negative ARF in this soft-plate system is found to be considerably broader than in stiff-plate systems [3, 4], with a full width at half maximum (FWHM) that is four times greater than that reported in Ref. [3]. It is further demonstrated by field distribution analysis that this negative ARF stems from the resonant vortex velocity field near the surface. The effects of varying the particle location and particle size on the ARF are also investigated.

Considering that acoustic transducers generate a broad frequency distribution rather than a single frequency, our soft-plate system is most suitable in practical applications where the trapping effect is less sensitive to the frequency spectra of the transducer. Moreover, because the broadband trapping forces guarantee sample insensitivity to the surface structure of the plate, simple samples can be used, thereby avoiding complicated sample fabrication routes. Finally, the broadband response of the negative ARF enables effective particle detection.

There are numerous prospective applications of such broadband negative ARFs,

including targeted drug delivery [24] and additive manufacturing [25], in which conventional sieves are typically used for separating smaller particles from the particle mixture. Acoustic sieves based on ARFs using our plate structure are expected to be capable of more complex particle manipulation, such as aligning, trapping, sorting, and transferring a large number of particles in a liquid. Plate-based ARFs could also potentially be used to design and control biological macromolecules because the acoustic forces are contactless and non-electromagnetic [26, 27].

## II. THEORY AND RESULTS

TABLE I. Material parameters.

|       | Density [kg/m$^3$] | Longitudinal velocity [m/s] | Transverse velocity [m/s] |
|-------|--------------------|-----------------------------|---------------------------|
| Water | 1000               | 1490                        |                           |
| Epoxy | 1800               | 2740                        | 1600                      |
| Brass | 8600               | 4400                        | 2100                      |

Referring to Fig. 1(a), the system under investigation is a water-immersed epoxy plate (yellow rectangle) of thickness $t$, patterned with a periodic brass grating (black squares) on the bottom. The period of the grating is the lattice spacing, denoted $d$; the side length of the brass strip is denoted $a$. A cylindrical brass particle (black circle) of radius $R$ is placed above the plate at a separation distance $\Delta y$ [inset of Fig. 1(a)] as measured from the bottom of particle to the plate, and at a distance $\Delta x$ from the center of the particle to the central axis perpendicular to the plate [corresponding to the $Y$-axis in Fig. 1(a)]. The material parameters are listed in Table I. The finite element method based on COMSOL Multiphysics software is employed to conduct numerical simulations, in which perfectly matched layers are used at the computational boundaries to prevent reflections. The overall size of the simulation domain is approximately $50d \times 10d$, corresponding to 50 grating periods. Unless otherwise specified, we set $R=0.2d$, $t=0.245d$, $a=0.21d$, $\Delta x=0.0d$, and $\Delta y=0.05d$. The frequency is normalized by $c_0/d$, where $c_0$ is the speed of sound in water.

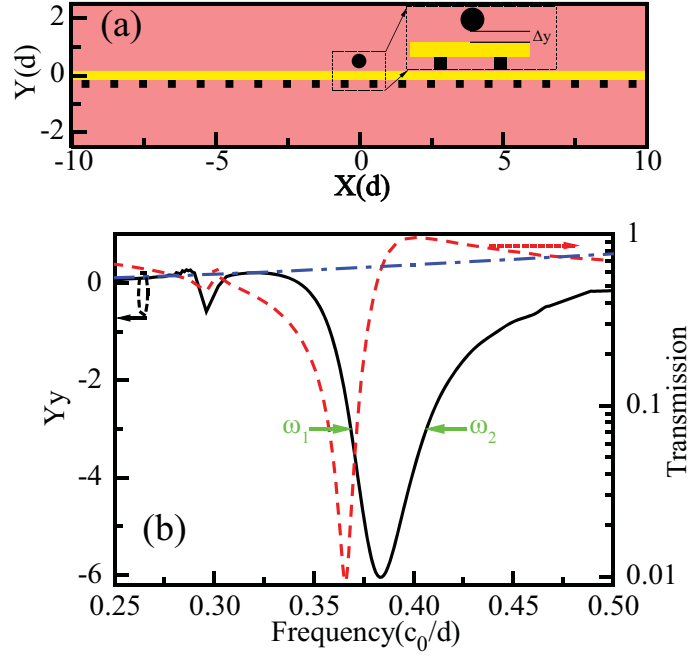

FIG. 1. (Color online) (a) Schematic view of the soft plate patterned with a periodic brass grating on the bottom of the plate. A cylindrical brass particle is placed above the structured plate. (b) The transmission (red dashed line; note the $\log_{10}$ scale) and dimensionless ARF function ($Y_y$; black solid line) representing the force exerted on the brass particle for the system employing a Gaussian beam (width $\approx 10d$) at normal incidence to the bottom of the plate. For comparison, the dimensionless ARF function representing the force exerted on the brass particle without the plate is also shown (blue dotted-dashed line).

The ARF acting on a cylindrical particle can be expressed as [9, 15, 17, 22, 23, 28, 29]

$$F = \oint \langle \mathbf{S} \rangle \cdot d\mathbf{A}, \quad (1)$$

where the differential area is $d\mathbf{A} = \mathbf{n} dA$, $\mathbf{n}$ is the normal force directed away from the particle, $A$ is the surface area of the particle at its equilibrium position, and $\langle \mathbf{S} \rangle$ is the time-averaged momentum-flux tensor, which is given by

$$\langle \mathbf{S} \rangle = \frac{\langle p^2 \rangle}{2\rho_0 c_0^2} \mathbf{I} - \frac{\rho_0 \langle |\mathbf{V}|^2 \rangle}{2} \mathbf{I} + \rho_0 \langle \mathrm{Re}(\mathbf{V}^*\mathbf{V}) \rangle. \quad (2)$$

Here, $\mathbf{V}$ and $p$ are the velocity and pressure fields, respectively, $\rho_0$ and $c_0$ are the mass density of, and sound speed in, the surrounding water, and $\mathbf{I}$ is a unit tensor. The dimensionless ARF function $Y_{x(y)} = F_{x(y)}/I_0$ is used to evaluate the force on the particle, where $I_0$ is defined as the incident acoustic energy density at the particle's position in the absence of a particle [3, 23, 28].

In Fig. 1(b), we show the transmission spectrum (red dashed line) and dimensionless ARF function ($Y_y$; black solid line) of the system with a Gaussian beam normally incident to the bottom of the plate. The width of the Gaussian beam is $10d$. For normal incidence, $Y_x$ becomes zero because of the symmetry of the system with respect to the $Y$-axis, and is therefore not displayed here. It can be seen from Fig. 1(b) that there is an abnormal transmission dip at a frequency of $0.368(c_0/d)$, which results from resonance excitation of antisymmetric Scholte waves in the plate [30–32]. Under such excitations, a collective resonance of the thin plate is induced, which further gives rise to a negative dynamic mass [31]. In addition, it is observed that $Y_y$ becomes negative over a broad range of frequencies centered on the frequency of the transmission dip, indicating that the brass particle is pulled toward the sound source at these frequencies. We note that similar negative ARFs have been observed for periodically structured brass plates, whereby the negative ARF spectra were centered around the resonance transmission peak of stiff-plate systems [3, 4]. For a quantitative comparison of the bandwidth of the pulling force in our soft-plate system and the stiff-plate system in Ref. [3], we calculated the FWHM values of the negative ARF spectra for both systems. The relative FWHM for the negative force of our soft-plate system is [$\omega_1$ and $\omega_2$ are marked by green arrows in Fig. 1(b)], which is found to be considerably larger (by a factor of about four) than that of the brass-plate system ($\Delta = 2.4\%$). Moreover, the dimensionless ARF function exerted on a brass particle immersed in a pure Gaussian beam without the plate is also shown for comparison [blue dotted-dashed line in Fig. 1(b)], and is characterized by uniform positive forces as expected.

## III. DISCUSSION

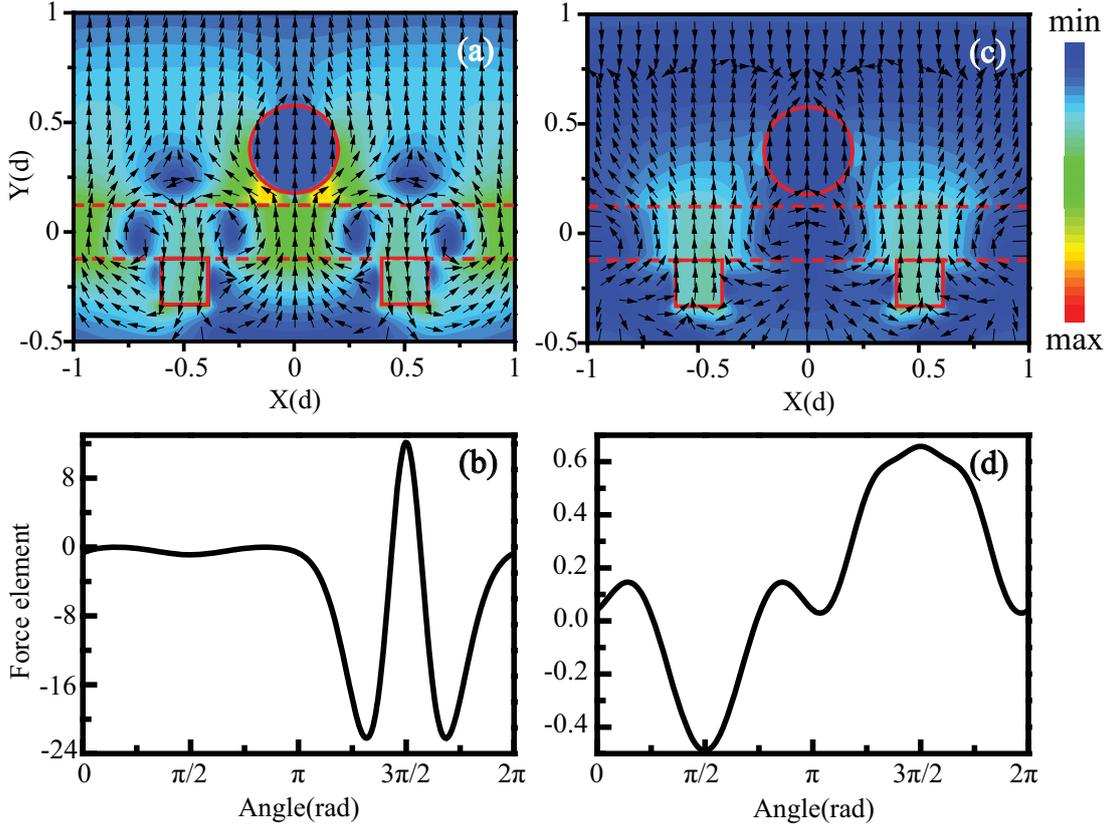

FIG. 2 (Color online) Velocity field distributions and *y*-direction force elements of our system at a resonance frequency of 0.383($c_0/d$) [(a) and (b)] and at a non-resonance frequency of 0.32($c_0/d$) [(c) and (d)]. The size and direction of the arrows in (a) and (c) indicate the magnitude and direction, respectively, of the velocity vector.

To further elucidate how the negative ARF is induced by antisymmetric Scholte waves under resonance excitation, we investigated the velocity field distributions and *y*-direction force elements across the particle for both negative and positive ARF frequencies. In Fig. 2(a), we show the velocity field distribution at a frequency of 0.383($c_0/d$) at which the maximum negative ARF is observed. Vortex and anti-vortex field patterns are clearly produced by the excitation of antisymmetric Scholte waves in the system; see Ref. [31] for details. We also calculated the *y*-direction force elements across the particle at 0.383($c_0/d$) [Fig. 2(b)]. The dimensionless ARF function over the whole particle can be obtained by integrating these force elements [3]. Here, the angular variable in Fig. 2(b) determines the location of the force element across the particle, which is measured with respect to the *X*-axis. We can see that the force elements are almost zero from 0 to $\pi$ across the particle. However, strong negative force elements are present at angles from $\pi$ to $4\pi/3$ and from $5\pi/3$ to $2\pi$. Although the particle undergoes strong pushing forces at angles from $4\pi/3$ to $5\pi/3$, after integrating all the force elements across the particle, the total force exerted on the particle is still negative. By analyzing the corresponding field patterns [Fig. 2(a)], we find that vortex and anti-vortex patterns exist in these regions, which lead to the negative forces on the particle. For comparison, we also calculated the field distributions and *y*-direction force elements across the brass particle at a normal frequency of 0.32($c_0/d$) without resonance. The results [Fig. 2(c) and (d)] show that there is no gradient field associated with these vortex and anti-vortex patterns, and the acoustic waves can transmit directly through the plate and exert a force on the particle. Accordingly, the force elements across the particle are almost zero in value, and the integration of the force elements is positive; thus, the particle is subjected to a pushing force.

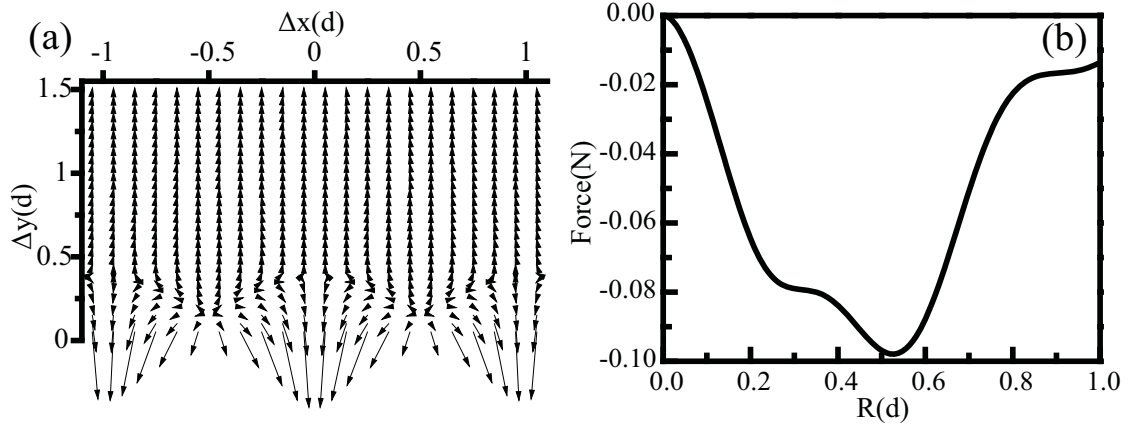

FIG. 3 (a) Distribution map of radiation forces exerted on the brass particle with position $\Delta x$ and $\Delta y$ for the soft-plate system under resonance. The particle radius is 0.2*d*. (b) Vertical component of the net force exerted on the particle as a function of particle radius with a constant separation of $\Delta y = 0.05d$.

Figure 3(a) shows the spatial distribution for total ARF of a particle with a radius of 0.2*d* at a frequency of 0.383($c_0/d$). The position of the particle ranges from −1.05*d* to 1.05*d* for $\Delta x$, and from 0.05*d* to 1.5*d* for $\Delta y$. The direction and length of the arrows in Fig. 3(a) indicate the ARF direction and intensity, respectively. Evidently, the particle undergoes pulling forces for values of $\Delta y$ smaller than 0.4*d*. In addition, the force becomes larger when the particle moves closer to the plate. This force distribution is consistent with the field distribution in Fig. 2. We also investigated the influence of particle radius on the ARF. In Fig. 3(b), the *y*-direction net force of the particle at position $\Delta x = 0.0d$ and $\Delta y = 0.05d$ is shown with different radii ($R$). A dip can be seen at $R = 0.5d$. From the radiation force map [Fig. 3(a)], we can conclude that when the total effective area of the particle subjected to a negative force (area immersed in the near-resonance field) becomes larger, the $y$-component of the pulling force becomes more negative. Hence, when the particle radius increases from zero to $0.5d$, the net pulling force increases [Fig. 3(b)]. However, a further increase in radius means the upper half of the particle is further away from the gradient resonance field, and this half would be subjected to a pushing (positive) force. Although the area of the particle in the near-resonance field increases, the interaction of these two competing forces results in a weakened negative force when the radius of the particle is further increased above the critical point ($R = 0.5d$).

As demonstrated in Section II, the negative ARF of our soft-plate system exhibits a broad bandwidth compared with stiff-plate systems. For both systems, negative ARFs arise from the gradients in the near-field of the vortex around the frequency range in which abnormal transmission (for stiff plates) and reflection (for soft plates) occur. For the stiff plate, the frequency range of abnormal transmission is usually narrower (and sometimes sharper) than that of abnormal reflection in a soft plate. This stems from differences in the material properties between stiff and soft plates. For a stiff plate, the intrinsic asymmetric modes are difficult to excite and the near-field vortex patterns do not appear unless the frequency of the excitation waves is very close to resonance; this leads to a narrow transmission spectrum [3,21]. In contrast, for our soft plate patterned with a periodic deep grating, as a combined result of the weak elastic modulus of the plate and the large mass density of the grating, intrinsic modes of this system can be excited even when the frequency of the external excitation is slightly off-resonance. This leads to a relatively broad reflection spectrum. In other words, near-field vortex patterns can be excited by a broadband acoustic wave impinging on our soft plate [31]. Therefore, our soft-plate system exhibits broadband characteristics not exhibited for stiff-plate systems.

It should be noted that the negative ARF dip is not consistent with the transmission dip in Fig. 1(b). As we have demonstrated, the collective resonance excitation of the antisymmetric, intrinsic Scholte surface waves in the plate leads to an abnormally enhanced reflection from the plate. Near-field vortex patterns in the velocity field distribution are produced when this resonance excitation occurs, which, in turn, give rise to the emergence of negative ARFs. Therefore, the emergence of such negative

ARFs should be accompanied by enhanced reflection for our soft-plate system. Nevertheless, the transmission dip is not guaranteed to be consistent with the ARF dip. The computed transmission, defined as the energy flux ratio, is proportional to the pressure field, $p$, in the far field. The expression for the *y*-component of the ARF acting on a cylindrical particle can be further reduced to [23]

$$F_y = \oint \left[ -\rho_0 \text{Re}(v_y^* v_x) dA_x + \left( \frac{\rho_0 |v_x|^2}{2} - \frac{|p|^2}{2\rho_0 c_0^2} - \frac{\rho_0 |v_y|^2}{2} \right) dA_y \right], \quad (3)$$

which indicates that both the velocity ($V$) and pressure field contribute to the ARF. In addition, the negative ARF experienced by the particle is obtained in the near-field of the plate, in which the pressure might be very different from that in the far-field. Therefore, at the frequency of the transmission dip, the combined contribution of the near-field $V$ and $p$ in Eq. (3) usually does not produce a maximal effect. This leads to the discrepancy in the frequency dips between the negative ARF and the transmission spectra because the former depends on the combined contribution of $V$ and $p$ in the near-field, whereas the latter is linked to $p$ of the far-field.

## IV. CONCLUSIONS

We have demonstrated that a broad and tunable negative ARF can be obtained in a system consisting of an epoxy plate patterned with a deep, periodic brass grating. The broadband feature of this negative force stems from collective excitations of antisymmetric coupled Scholte waves in the soft plate. It is further demonstrated that this negative ARF is caused by the gradient force induced by the gradient vortex velocity field near the surface. The effects of particle location and size on the ARF were also investigated in detail. Negative ARFs have potential use in applications based on acoustic trapping and manipulation of particles.

## ACKNOWLEDGMENTS

This work is supported by the National Natural Science Foundation of China (Grant No. 11104113, 11304119, 11464012 and 11264011), the Natural Science Foundation of Hunan Province, China (Grant No. 13JJ6059), and the Natural Science Foundation of Education Department of Hunan Province, China (Grant No. 13B091, 13A077 and 13C750).